\documentclass{article}

\usepackage{microtype}
\usepackage{graphicx}
\usepackage{subfigure}
\usepackage{booktabs} 

\usepackage[final]{hyperref}

\usepackage[accepted]{icml2019}

\usepackage{graphicx}
\usepackage{amsmath}
\graphicspath{{figures/}}
\usepackage{bm} 
\usepackage{amssymb}
\usepackage{amsthm}

\icmltitlerunning{Submission and Formatting Instructions for ICML 2019}

\begin{document}

\twocolumn[
\icmltitle{Optimistic Bull or Pessimistic Bear: Adaptive Deep Reinforcement Learning for Stock Portfolio Allocation}



\icmlsetsymbol{equal}{*}

\begin{icmlauthorlist}
\icmlauthor{Xinyi Li}{equal,to}
\icmlauthor{Yinchuan Li}{equal,goo,ed}
\icmlauthor{Yuancheng Zhan}{si}
\icmlauthor{Xiao-Yang Liu}{goo}
\end{icmlauthorlist}

\icmlaffiliation{to}{Department of Statistics, Columbia University, New York, US}
\icmlaffiliation{goo}{Electrical Engineering, Columbia University, New York, US}
\icmlaffiliation{ed}{School of Information and Electronics, Beijing Institute of Technology, Beijing, China}
\icmlaffiliation{si}{Electronic Engineering and Information Science, University of Science and Technology of China, Hefei, China}

\icmlcorrespondingauthor{Yinchuan Li}{yl3923@columbia.edu}

\icmlkeywords{Machine Learning, ICML}

\vskip 0.3in
]



\printAffiliationsAndNotice{\icmlEqualContribution} 

\begin{abstract}

Portfolio allocation is crucial for investment companies. However, getting the best strategy in a complex and dynamic stock market is challenging. In this paper, we propose a novel Adaptive Deep Deterministic Reinforcement Learning scheme (Adaptive DDPG) for the portfolio allocation task, which incorporates optimistic or pessimistic deep reinforcement learning that is reflected in the influence from prediction errors. Dow Jones $30$ component stocks are selected as our trading stocks and their daily prices are used as the training and testing data. We train the Adaptive DDPG agent and obtain a trading strategy. The Adaptive DDPG's performance is compared with the vanilla DDPG, Dow Jones Industrial Average index and the traditional min-variance and mean-variance portfolio allocation strategies. Adaptive DDPG outperforms the baselines in terms of the investment return and the Sharpe ratio.

\end{abstract}

\section{Introduction}
\label{introduction}

Portfolio allocation plays an important role in the financial market, which is fundamental and important for investment companies and quantitative analysts. The famous economist, Harry Markowitz received the 1990 Nobel Memorial Price in Economic Sciences for his pioneering theoretical contributions to financial economics and corporate finance. His innovative work laid the foundation for Modern Portfolio Theory (MPT) \cite{sharpe1970portfolio}, i.e., construction of a portfolio to maximize the expected return while minimizing the investment risk.
Portfolio theory studies how ``rational investors'' optimize their portfolios. A rational investor may maximize the expected return at a given level of expected risk or minimize the expected risk at a given expected level of return. The essence of investment is a tradeoff between the profits and risks of uncertainty.

 The target of portfolio allocation is that either maximizing the Sharpe ratio (average return minus the risk-free return divided by the standard deviation) or minimizing the risk for a range of returns. Portfolio theory uses the mean-variance to characterize these two key factors. The mean value refers to the weighted average of expected return, and the weight is the allocated fraction of investment. The variance refers to the variance of the expected return of the portfolio. We refer to the standard deviation of expected return as the volatility, which portrays the risk of the portfolio. Traditional approach is performed in two steps as described in \cite{markowitz1952portfolio}. First, the expected returns of the stocks and the covariance matrix of the stock prices are computed. The trading strategy is then extracted by following the portfolio allocation. 
 
 However, the challenge of the traditional portfolio allocation is that the approach can be very complicated to implement if the manager wants to revise the decisions made at each time step and take, for example, transaction cost into consideration. 
The financial market is complex and being influenced by immense factors such as macroeconomic, traders' expectation and risk aversion.  Investment companies are eager to optimize allocation of capital and thus maximize performance, for example, get a higher return while the risk is as small as possible. 
 Despite traditional portfolio allocation's momentous theoretical importance, many people who criticize MPT believe that the basic assumptions and models of their financial markets are inconsistent with the real world in numerous respects \cite{mangram2013simplified}. 
 In general, some key criticisms include: investor ‘irrationality’, high risk is accompanied by high returns, no taxes or transaction costs, the market is efficient, investment independence, unlimited access to capital and investors may get perfect information. 

In this paper, we apply the optimistic and pessimistic reinforcement learning for stock portfolio allocation. The existing approach to solve the stock trading problem is to model it as a Markov Decision Process (MDP) and use dynamic programming (DP) to obtain the optimum strategy. 
But the DP algorithms have limited practical usages because they assume a perfect model and they are also compute-intensive.  
The scalability of DP is limited due to the state space explosion when dealing with the stock market. Reinforcement learning has no such restrictions. Deep reinforcement learning methods use a function approximator and a stochastic approximation to calculate a relevant expectation, which can be applied to problems with a large continuous state space \cite{neuneier1998enhancing}. Then, we explore the Deep Deterministic Policy Gradient (DDPG) \cite{lillicrap2015continuous}\cite{Liu2019NIPS}\cite{Wenhang2019ICML}, to find the best trading strategy in the complex and dynamic stock market. 

We adopt the DDGP algorithm that consists of three key components: (i) actor-critic framework \cite{konda2000actor} that models large state and action spaces; (ii) target network that stabilizes the training process \cite{mnih2015human}; (iii) experience replay that removes the correlations between samples and increases the usage of transition data. The efficiency of DDPG algorithm is demonstrated by achieving higher return than the traditional portfolio allocation method and the Dow Jones Industrial Average index. 

Furthermore, the proposed deep reinforcement learning scheme takes into account the impact of the market index, which is very meaningful in practice. Since the machine learning based methods are more objective and more quantitative than the trader's decision based on the market. Theoretical studies of behavioral finance have shown that environment can influence investment decisions. Investors may optimistic and pessimistic, as behavioral finance has asserted \cite {li2014effect}. In general, the bear market occurred during a recession or depression, when pessimism occurred. When the price of securities rises faster than the overall average interest rate, there will be a bull market. Bull market is accompanied by economic growth and investor optimism. 
We hence proposed a modified Rescorla-Wanger model~\cite{lefebvre2017behavioural}, which can learn differently from positive and negative environment, which can calculate reward of choosing different options (buy, hold and sell). The model can adjust the amplitude of changes according to the sign of prediction errors. For example, when the prediction error is positive (actual reward is better than the expected reward), then the learning rate would adjust the amplitude from one trial to the next. The model can distinguish between good and bad environment feedback. Furthermore, we propose the optimistic and pessimistic reinforcement learning model that are applicable to both the bear market and the bull market, respectively.

The reminder of this paper is organized as follows. Section 2 contains background of stock portfolio allocation. In Section 3, we drive and specify the main DDPG algorithm and  optimistic-pessimistic deep reinforcement learning. Section 4 describes data preprocessing, experimental setup and presents the performance of Adaptive DDPG model. Section 5 gives our conclusions.

\section{Problem Formualtion for Stock Portfolio Allocation}

In this section, we introduce the portfolio allocation model, the impact of the market environment on stocks, and the relevant reinforcement learning method that can be applied to the stock trading strategy.


\subsection{Portfolio Allocation}

\begin{figure*}[!tb]
	\centering
	
	{\includegraphics[width=3.0in]{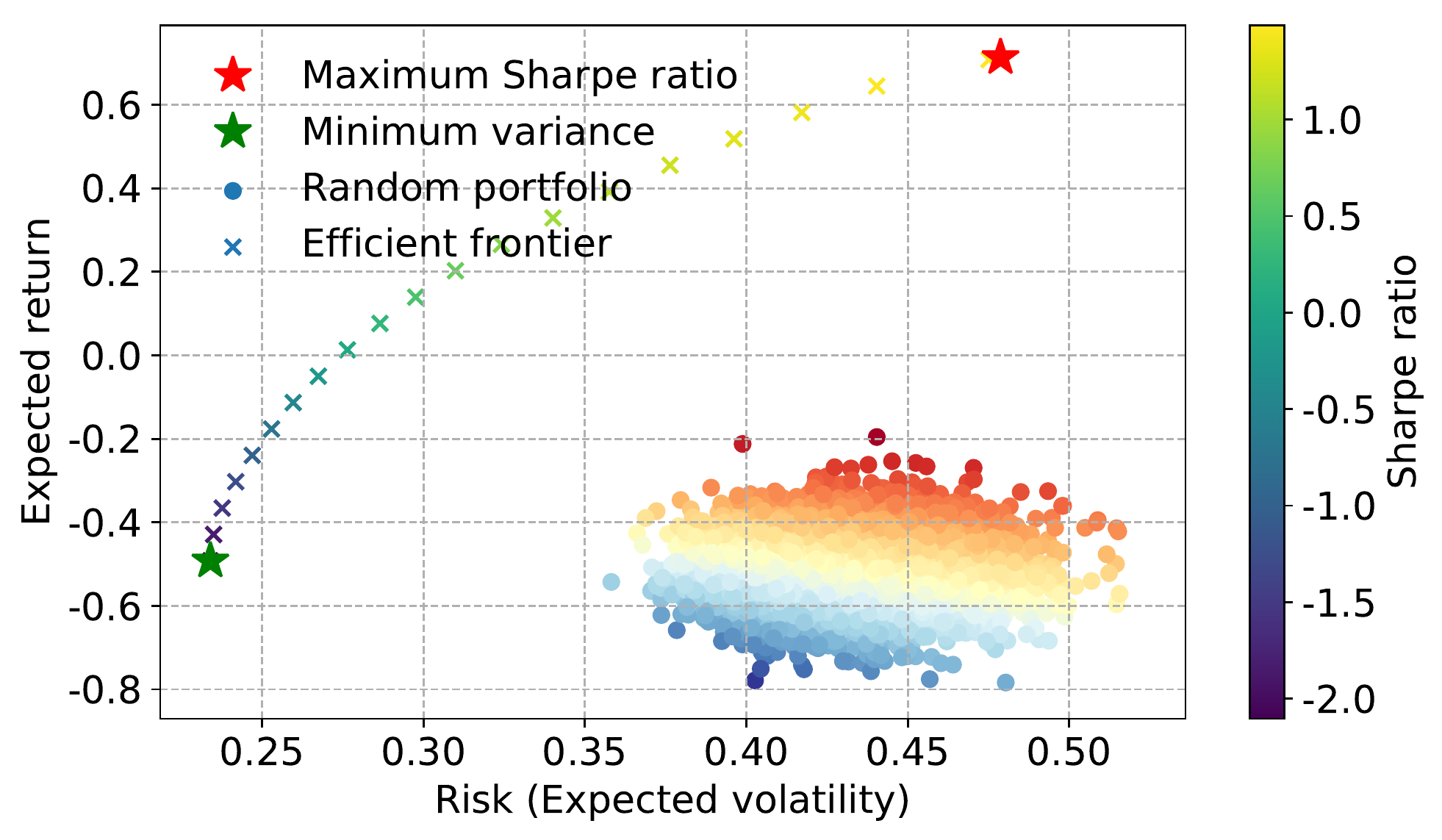}}
	{\includegraphics[width=3.0in]{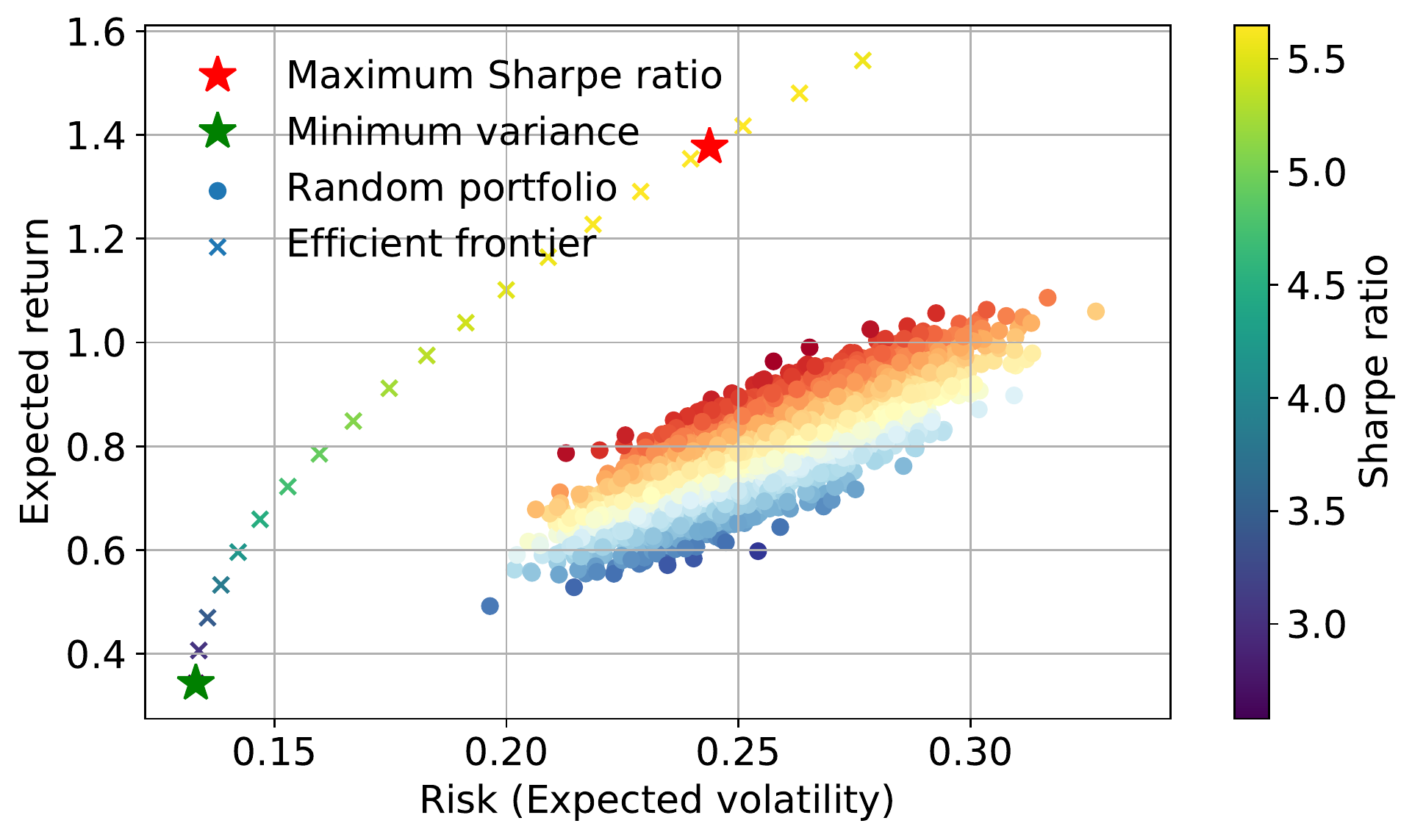}}
	
	\caption{Portfolio allocation for: (a) 01/05/2009-03/31/2009 (during financial crisis); (b) 04/01/2009-06/25/2009 (after financial crisis).}
	\label{figure:portfolio1}
\end{figure*}

	


\subsubsection{Portfolio Allocation}

The portfolio theory discussed in this paper is narrowly defined. In the developed securities market, Markowitz's portfolio theory has proven to be effective in practice and is widely used in portfolio selection and asset allocation. 

The theory contains two important parts: 1) the mean-variance analysis method; 2) and the portfolio efficient frontier. Specifically, the optimized investment portfolio is depicted in a two-dimensional plane with the volatility as the abscissa and the return as the ordinate, forming a curve. There is a point on this curve with the lowest volatility, called {\em the min-variance point (MVP)}. The portion of this curve above the min-variance point is the well-known Markowitz portfolio effective boundary, and the corresponding portfolio is called the effective portfolio. The portfolio's efficient frontier is a monotonically increasing convex curve. However, because the traditional effective market hypothesis cannot explain market anomalies, portfolio theory is challenged by the behavioral finance theory.

\subsubsection{Traditional Methods}

 We provide two base portfolio allocation methods for investors with different risk levels. The first one is using mean-variance optimization to allocate the stocks, this method is suitable for the investors who prefer a higher sharp ratio. The second one is min-variance portfolio with the lowest possible risk for investors. 
 
\textbf{Mean variance method:}

Markowitz's work~\cite{sharpe1970portfolio} show that it is not a security's own risk that is important to an investor, but rather the contribution the security makes to the variance of the entire portfolio. This follows from the relation between the variance of the return of a portfolio  ($\sigma_p^2$)  and the variance of the return of its constituent securities  ($\sigma_i^2$, $i=1,2,…,m$). Calculate the annualized rate of return $\mu_p$ and covariance matrix $\Sigma_p$ as follows:
\begin{align}
\label{Mean}
\mu_p&=E(r_p)= \sum_{i=1}^{m}w_i E(r_i)=W^T\mu, \\
\sigma_p^2&=\sum_{i,j=1}^{m}w_iw_j\sigma_{i,j}=W^T\Sigma_p W,
\end{align}
where $m$ is the number of stocks, $w_i$ is the weight of $i$-th stock, which is  the portfolio percentage. $\sigma_{ij}$ is the covariance of the returns of stock $i$ and $j$. Then we set constraints for the mean-variance:
\begin{align}
\label{con_mv1}
w_i \in [0,0.2],i=1,...,m;~~\sum_{i=1}^{m}w_i =1,
\end{align}
where 0 and 0.2 is the lower bound and upper bound of allocation weight. Our objective function is to find the allocation that makes the highest sharp ratio (the portfolio of red stars in Figure~\ref{figure:portfolio1}).

\textbf{Min-variance method:}

The min-variance method is similar with the mean variance method, except the objective function is replaced by finding the portfolio with the smallest variance (the portfolio of green stars in Figure~\ref{figure:portfolio1}).

Therefore, the decision to hold securities should not be made solely by comparing its expected returns and variance with other stocks, but depend on the other stocks the investor wants to hold. Stocks should be properly evaluated as a group instead of in isolation.

\subsubsection{Limitation of Modern Portfolio Theory}
As we mentioned in the Introduction, none of these MPT's assumptions are entirely true and the assumptions compromised MPT. The key limitations of the MPT are as follows:
\begin{itemize}
\item \textbf{Assumption 1:  investors are rational.}

MPT assumes that the investor is rational and seeks to maximize returns while minimizing risk. This contradicts the observations of market participants who were involved in the “herd behavior” investment activity \cite{morien12travis}. For example, investors often choose “hot” stocks, and because of speculative excessive behavior, markets often experience prosperity or depression. Large stock market trends often begin and end with periods of frenzied buying (bubbles) or selling (crashes). These herding behavior that is irrational and driven by emotion-greed in the bubbles, fear in the crashes. Even if herd behaviors might be rare, this has important consequences for a whole range of real markets. 






\item \textbf{Assumption 2:  the market is efficient.}

Markowitz theoretical assumes that the market is fully valid \cite{markowitz1952portfolio}. In contrast, it does not consider potential market failures such as information asymmetry. A hundred years of  ``prosperity'', ``depression'', ``bubble'' and ``financial crisis'' indicate that the market is far from efficient. Using the market index to do the  portfolio allocation can overcome the market's ineffective problems to some extent. 





\item \textbf{Assumption 3:  investment are independent.}

MPT assumes that securities whose individual performance is independent of other securities are selected.  However, during market pressures and extreme uncertainty, seemingly independent investments actually show relevance. Market history has proven that there is no such tool \cite{mcclure2010modern}. Stocks and markets cannot be separated, and the influence of market environment on stocks must be considered.
\end{itemize}

MPT seeks to maximize risk-adjusted returns while ignoring environmental, personal, strategic or social factors. The historical `expected value' assumptions often fail to consider that the updated environment does not exist during historical data.

\subsection{Market Environment}





The market price of a stock is determined by the value of the stock, but at the same time it is affected by many other factors. Generally, the factors affecting the stock market price mainly include the following two aspects:
\begin{itemize}
\item Macroeconomic factors: The impact of the macroeconomic environment and its changes on the stock market price, including the regular factors such as the cyclical fluctuations of macroeconomic operations and the policy factors such as the economic policies implemented by the government \cite{flannery2002macroeconomic}. The stock market is an important part of the entire financial market system.  Therefore, stock prices in the stock market will naturally change with the macroeconomic conditions. For example, in general, stock prices fluctuate with the rise and fall of gross national products.
\item Market factors:
Various stock market operations may affect the stock market price. For example, bullish and bearish, short and short selling, chasing and killing \cite{chang2009integrating}. Generally speaking, if in the bull market, investors intend to be more aggressive, and the stock price will rise. Conversely, if the short-selling behavior prevails and the investor is overwhelmed, the stock price tends to fall. Since various stock market operations are mainly short-term behaviors, the impact of market factors on stock market prices has a clearly short-term nature.
\end{itemize}

To illustrate the impacts of the market's overall environment on the portfolio, we compared the portfolio strategy for three months before and after the end of the financial crisis. The Stock Pool is Dow Jones 30. Figure~\ref{figure:portfolio1}(a) plots the portfolio allocation during the financial crisis and Figure~\ref{figure:portfolio1}(b) plots the portfolio allocation after the financial crisis. Comparing these two figures, we find that in the financial crisis, the investment portfolio generally shows a low return and high volatility state; after the financial crisis, the economy begins to recover, and the portfolio presents relatively low volatility and higher returns.

An emerging area of applied reinforcement learning is stock market trading, in which a trader's behavior is similar to an agent because buying and selling are particular actions. Reward is that stock changes the state of the trader by generating profit or loss.

\subsection{ Markov Decision Process Formulation}
MDP is particularly important for reinforcement learning. A particular MDP is defined by its state $s$ and action set $a$ and the one-step dynamics of the environment. Given any state and action, the probability of each possible next state $s^{\prime}$ is as follows
\begin{align}
\label{MDP}
\mathcal{P}_{s s^{\prime}}^{a}=\operatorname{Pr}\left\{s_{t+1}=s^{\prime} | s_{t}=s, a_{t}=a\right\}.
\end{align}
These quantities are called transition probabilities. Similarly, the expected value of the next reward is
\begin{align}
\label{MDP2}
\mathcal{R}_{s s^{\prime}}^{a}=\mathbb{E}\left\{r_{t+1} | s_{t}=s, a_{t}=a, s_{t+1}=s^{\prime}\right\}.
\end{align}
We assume that the environment is a finite MDP. The quantities $\mathcal{P}_{s s^{\prime}}$ and $\mathcal{R}_{s s^{\prime}}^{a}$ specify the most important aspects of limited MDP dynamics.

\subsection{Relevant Reinforcement learning method}
As mentioned in the Introduction, DDPG is mainly developed from: PG (Policy Gradient)$\rightarrow$DPG (Deterministic Policy Gradient)$\rightarrow$DDPG (Deep Deterministic Policy Gradient). Next, we will introduce this evolution process to give reasons for adopting DDGP to find the optimal trading strategy in the complex and dynamic stock market. 

\section{Adaptive Deep Reinforcement Learning}

\label{reinforcement}
We model the stock trading process as a MDP. We then formulate our trading goal as a maximization problem. The portfolio allocation task is formalized as MDP under the following assumptions:

\begin{enumerate}
\item Investors can trade at each time step along a continuous infinite time horizon.
\item A single investor's trading cannot influence the market.
\item There are only two kinds of assets (stocks and risk-free asset) for investing the capital.
\item The investor invests the total amount, which means investors has no risk aversion.
\end{enumerate}

\subsection{Basic Idea}

Training intelligent agents for automatic financial assertion transactions is a long-standing topic that has been widely discussed in modern artificial intelligence \cite{saad1998comparative}. In essence, the transaction process is well described as an online decision problem, which involves two key steps: market environment and best action execution. Due to the lack of supervision information, dynamic decision making is more challenging. Therefore, it requires agents to explore the unknown stock market environment themselves and make the right decisions at the same time. 

Compared to traditional reinforcement learning tasks, algorithmic trading is much more difficult due to the following challenges: 1) the challenge stems from the difficulty of summarizing and representing the financial environment; 2) financial data contains a lot of noise, jumping and moving, resulting in a very unstable time series. 

In summary, the reinforcement learning trains the agent to interact with an environment to obtain the maximum total reward. This bonus value is generally associated with the mission target defined by the agent.

\subsubsection{Basic concepts related to DDPG}

\begin{figure}[!tb]
	\centering
	{\includegraphics[width=3.0in]{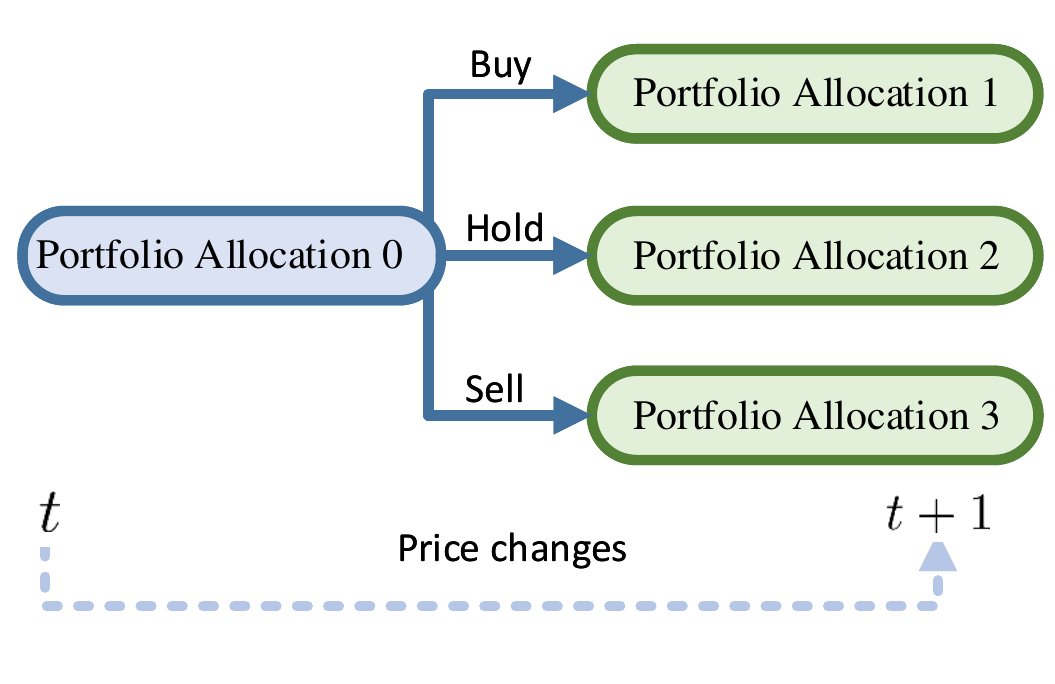}}
	\caption{Portfolio allocation 0 is the starting state at time $t$, three actions (buy, hold and sell) lead to three possible portfolios at time $t+1$ after price changes.}
	\label{figure:buy}
\end{figure}

In order to address the above problems and considering the stochastic and interactive nature of the trading market, we model the stock trading process as a Markov Decision Process (MDP) as shown in Figure~\ref{figure:buy}, which is specified as six parts: state, action, reward, discounted future reward, policy and action-value. 

State $s= [p,w,b]$: $s$ is the state of the environment. These states are generated based on the agent's behavior strategy. $s$ is a set that includes the information of the prices of stocks $p\in \mathbb{R}_+^D$, the weight of holdings of stocks $w=(w_1,w_2,...,w_D)^T$; $w_i\in [0,1],~ i=1,2,...,D;~ \sum_{i=1}^{D}w_i=1$, and the remaining balance $b\in \mathbb{R}_+$, where $D$ is the number of stocks that we consider in the market and $\mathbb{Z}_+$ denotes non-negative integer numbers.

Action $a$: a set of actions on all $D$ stocks. The available actions of each stock include selling, buying, and holding, which result in decreasing, increasing, and no change of the weight of holdings $w$, respectively. DDPG is a learning method for continuous behavior. In our model, the action is continuous, because our weights are continuously changing.

Reward $r(s,a,s')$: 
the change of the portfolio value when action $a$ is taken at state $s$ and arriving at the new state $s'$. The portfolio value is the sum of the equities in all held stocks $p^T w$ and balance $b$. We also called $r(s,a,s')$ as the single-step reward value, returned by the environment after the action $a$ is executed in the states. The above relationship can be represented by a state transition in Figure~\ref{figure:state}:

\begin{figure}[!htb]
	\centering
	{\includegraphics[width=2.3in]{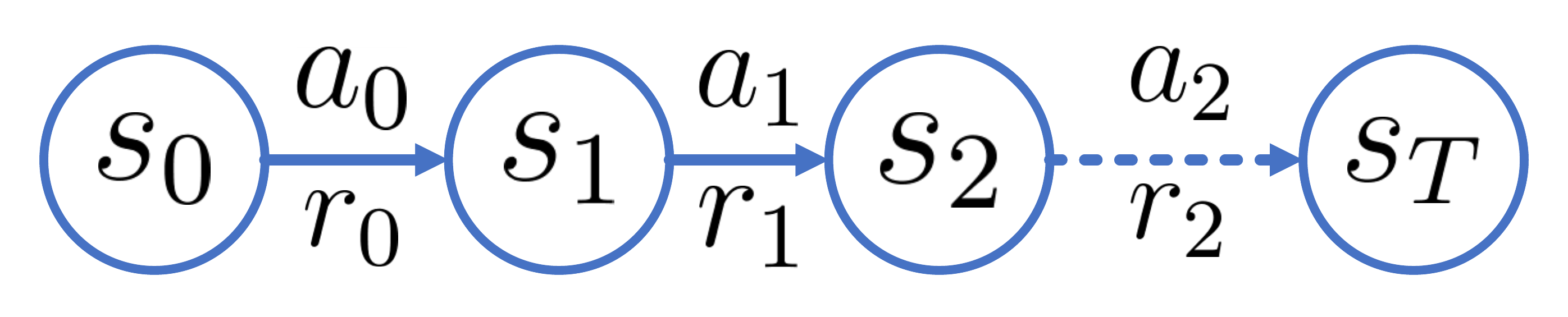}}
	\caption{State transition diagram.}
	\label{figure:state}
\end{figure}

Discounted future reward $R_t$: is the weighted sum of the reward values obtained for all actions from the current state to a future state.
\begin{align}
\label{discount}
    R_{t}=\sum_{i=t}^{T} \gamma^{i-t} r\left(s_{i}, a_{i},s_{i+1}\right),
\end{align}
where $\gamma$ is called discounted rate, $\gamma\in[0,1]$, usually $\gamma=0.99$.

Policy $\pi(s)$: the trading strategy of stocks at state $s$. It is essentially the probability distribution of $a$ at state $s$. To be more specific, given a state, the decision policy will calculate the next action to take.

Action-value function $Q_\pi(s,a)$: the expected reward achieved by action $a$ at state $s$ following policy $\pi$.

 \subsubsection{Framework of stock market dynamics}
 
One of the solutions to the stock trading problem is to model it as a MDP and use dynamic programming (DP) to solve the optimal strategy. However,  DP only solves problems with small discrete state spaces. 

Driven by these challenges, we explore the deep reinforcement learning algorithm DDPG \cite{lillicrap2015continuous}\cite{Liu2019NIPS}\cite{Wenhang2019ICML}, to find the best trading strategies in complex dynamic stock markets. Most reinforcement learning algorithms boil down to just three main steps: infer, perform, and learn. During the first step, the algorithm selects the best action $a$ given a state $s$ using the knowledge it has so far. Next, it performs the action to find out the reward $r$ as well as the next state $s$. Then, it improves its understanding of the world using the newly acquired knowledge. We would describe the framework of stock market dynamics as follows. We use subscript to denote time $t$, and the available actions on stock $d$ are:
\begin{itemize}
    \item Selling: $k$ ($k\in [0,w_d]$, where $d = 1,...,D$) weight of shares can be sold from the current holdings, where $k$ must be a weight. In this case, $w_{t+1}=w_t-k$.
    \item Holding: $k=0$ and it leads to no change in $w_t$.
    \item Buying: $k$ weight of shares can be bought and it leads to $w_{t+1}=w_t+k$. In this case $a_t[d]=-k$ is  negative weight.
\end{itemize}
It should be noted that all bought stocks should not result in a negative balance on the portfolio value. That is, without loss of generality we assume that selling orders are made on the first $d_1$ stocks and the buying orders are made on the last $d_2$ ones, and that $a_t$ should satisfy $p_t[1:d_1]^T a_t[1:d_1]+b_t+ p_t[D-d_2:D]^T a_t[D-d_2:D]\ge 0$. The remaining balance is updated as $b_{t+1}=b_t+p_t^T a_t$.
As defined above, the portfolio value consists of the balance and sum of the equities in all held stocks. 
At time $t$, an action is taken, and based on the executed action and the updates of stock prices, the portfolio values change from ``portfolio value 0'' to ``portfolio value 1'', ``portfolio value 2'', or ``portfolio value 3'' at time $(t+1)$. 





According to Bellman Equation, the expected reward of taking action $a_t$ is calculated by taking the expectation of the rewards $r(s_t,a_t,s_{t+1})$, plus the expected reward in the next state $s_{t+1}$. Based on the assumption that the returns are discounted by a factor of $\gamma$, we have
\begin{align}
    \label{decom}
    Q_\pi(s_{t},a_t) = &~\mathbb{E}_{s_{t+1}} \{r(s_t,a_t,s_{t+1})  \nonumber\\
    &+ \gamma \mathbb{E}_{a_{t+1} \sim \pi(s_{t+1})}[ Q_{\pi} (s_{t+1},a_{t+1})] \}.
\end{align}
The above $Q$ function is the action-value function, defined in the state $s_t$, after taking the action $a_t$.


The goal is to design a trading strategy that maximizes the investment return at a target time $t_f$ in the future, i.e., $p_{t_f}^T w_t+b_{t_f}$, which is also equivalent to $\sum_{t=1}^{t_f-1}r(s_t,a_t,s_{t+1})$. Due to the Markov property of the model, the problem can be boiled down to optimizing the policy that maximizes the function $Q_\pi (s_t,a_t)$. This problem is very hard because the action-value function is unknown to the policy maker and has to be learned via interacting with the environment. Hence in this paper, we employ the optimistic \& pessimistic deep reinforcement learning approach to solve this problem.

\subsection{Optimized Model Incorporates Market Environment}


In order to incorporate the market information into the deep reinforcement learning, we propose an effective method to quantitatively analyze the mechanism of stock information penetration.

\subsubsection{Modified Rescorla-Wanger model}
The computational part include a Rescorla-Wanger model (also called Q-learning, in the following referred as RW model). Base on the RW model, we use a modified model which learns differently from positive and negative environment emotional news. Positive and negative prediction errors representing positive and negative environment emotional news respectively (refer to as RW$\pm$). For each state, Q-values represent the expected reward by taking a specific action in a given market environment. Considering that we have three actions buying, holding and selling, the model estimates the expected values of buying, holding and selling options, on the basis of sequence actions and outcomes. The initial Q-value is set as 0 before learning. In each step $t$, the value of the options (buy, hold and sell) is updated according to the rules, as follows:
\begin{align}
\label{Qbuy}
    Q_{\pi}(s_{t+1},a_{t+1})=Q_{\pi}(s_t,a_t)+\alpha\delta(t).
\end{align}
where $\alpha$ is the learning rate, which is a scaling parameter that adjusts the magnitude of the change from one trial to the next, and $\delta(t)$ is the prediction error (we also define as environment emotional news), calculated as follows:
\begin{align}
    \delta(t)=r(s_t,a_t,s_{t+1})-Q_{\pi}(s_t,a_t),
\end{align}
which is the difference between the expected reward of $Q_{\pi}(s_t,a_t)$ and the actual reward $r(s_t,a_t,s_{t+1})$. Following this rule, the option value is increased if the result is better than expected, while the option value is decreased in the opposite case, and the amplitude of the update is similar after the positive, neutral and negative prediction errors.

\subsubsection{Update rule }
The update rule of the modified Q-learning algorithm (RW$\pm$) is given by \cite{lefebvre2017behavioural}
\begin{align}
\label{Qal_buy}
    Q_{\pi}(s_{t+1},a_{t+1})=Q_{\pi}(s_t,a_t)+ \left\{ \begin{array}{cc}{\alpha^{+} \delta(t)} & {\text { if } \delta(t)>0,} \\ {\alpha^{-} \delta(t)} & {\text { if } \delta(t)<0.} \end{array} \right.
\end{align}
When the prediction error is positive, which means the actual reward $r(s_t,a_t,s_{t+1})$ is better than the expected reward $Q_\pi(s_t,a_t)$, the learning rate $\alpha^+$ adjusts the amplitude of the change from one trial to the next, and vice versa. 
Therefore, the RW$\pm$ model allows the amplitude of the update to be different, a following sequential positive (‘good  environment emotional news’) and negative (‘bad  environment emotional news’) prediction errors. The RW$\pm$ model also allows us to consider individual differences in the way which learn from positive and negative experiences.








Furthermore, given the Q-values, the associated policy of selecting each options (buy, hold and sell) was estimated by implementing the softmax rule as follows:
\begin{align}
\label{soft}
   \pi(s_t)=e^{\left(Q_{\pi}(s_t,a_t) \beta\right)} /\left(e^{\left(Q_{\pi}(s_t,a_t) \beta\right)}+e^{\left(Q_{\pi}(s_t,a_t) \beta\right)}\right),
\end{align}
where $\beta$ is a scaling parameter that adjusts the stochasticity of decision-making, which is used to control the exploration–exploitation trade-off. \eqref{soft} is a standard stochastic decision rule that calculates the probability of selecting one of a set of options based on the associated value. 

\subsection{Adaptive DDPG}

\begin{figure}[!tb]
	{\includegraphics[width=3.2in]{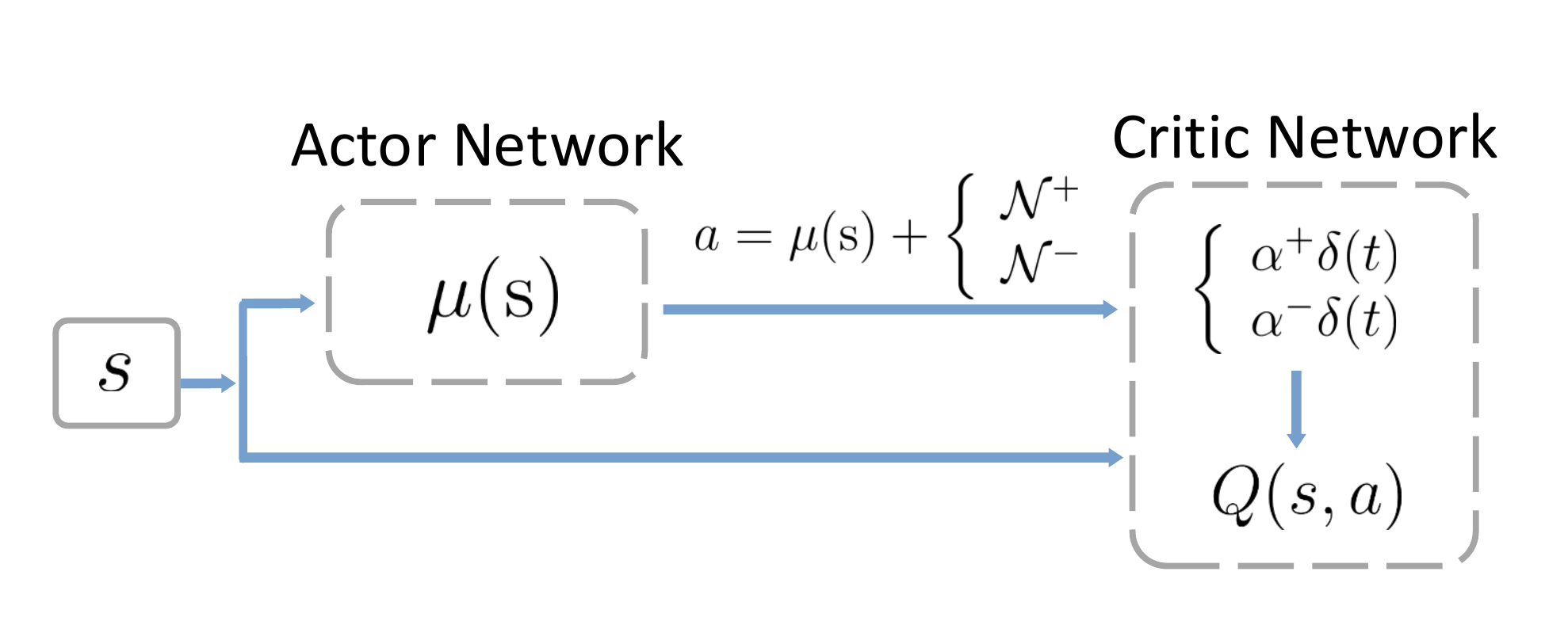}}
	\caption{Learning network architecture.}
	\label{figure:network}
\end{figure}

We use the adaptive DDPG algorithm to maximize return on investment. DDPG is an improved version of the Deterministic Policy Gradient (DPG) algorithm and DPG is based on Policy Gradient (PG) improvements. As for the DDPG, Q-learning uses greedy action $a_{t+1}$ to maximize $Q\left(s_{t+1}, a_{t+1}\right)$ for state $s_{t+1}$ as follow
\begin{align}
    \label{greedy}
    &Q_{\pi}\left(s_{t}, a_{t}\right)  \nonumber \\
    =&~\mathbb{E}_{s_{t+1}}\left[r\left(s_{t}, a_{t}, s_{t+1}\right)+\gamma \max _{a_{t+1}} Q\left(s_{t+1}, a_{t+1}\right)\right].
\end{align}
 As shown in Figure~\ref{figure:network}, the adaptive DDPG includes an actor network and a critic network. The actor network $\mu\left(s | \theta^{\mu}\right)$ maps state to actions, and after the prediction error $\delta(t)$ is available, the critic network then updates $Q\left(s, a | \theta^{Q}\right)$ according to the prediction error $\delta(t)$ and the learning rate $\alpha^+$ (or $\alpha^-$), where $\theta^\mu$ is the set of actor network parameters and $\theta^Q$ is the set of critic network parameters. $\mathcal{N}^{+}$ and $\mathcal{N}^{-}$ are the random processes corresponding to the positive and negative environment respectively, which are used to add noise to the output of the actor network to explore actions.
 
 
 Similar to DDPG, our model uses the experience replay buffer $R$ to store transitions. The adaptive DDPG agent updates $Q(s,a)$ according to the prediction error $\delta(t)$ and takes an action $a_t$ on $s_t$, and then receives a reward based on $s_{t+1}$. Then calculate $y_i = r_i+\gamma Q' (s_{i+1}, \mu' (s_{i+1}|\theta^{\mu'},\theta^{Q'})), i=1,...,N$. The transition $(s_t,a_t,s_{t+1},r_t)$ is then stored in replay buffer $R$. After $N$ sample transitions are drawn from $R$, we update the critic network by minimizing the expected difference $L(\theta^Q)$ between outputs of the target critic network $Q'$ and the critic network $Q$.

After the critic network and the actor network are updated by the transitions from the experience buffer, the target actor network and the target critic network are updated as follows:
\begin{align}
    \theta^{Q'} &\leftarrow \tau \theta^Q + (1-\tau)\theta^{Q'},
    \\
    \theta^{\mu'} &\leftarrow \tau \theta^\mu + (1-\tau)\theta^{\mu'},
\end{align}

where $\tau$ denotes learning rate.

\section{Performance Evaluation}
We evaluate the performance of the adaptive DDPG algorithm in this section. The results show that the adaptive DDPG agent achieves higher return than the vanilla DDPG, Dow Jones Industrial Average (DJIA) and the traditional portfolio allocation strategies.
\subsection{Data Preprocessing and Comparison Methods}
We selected the constituent stocks of the Dow Jones index as our stock pool. The time span of the data (daily prices) is from 01/01/2001 to 09/30/2018. The dataset is downloaded from Compustat database accessed through Wharton Research Data Services (WRDS). The dataset from 01/01/2001 to 12/30/2013 (including 3268 trading days) is used as the training data, and the remaining dataset (from 01/02/2014 to 10/02/2018 including 1190 trading days) is used as the testing data. We train our agent on training data and test the agent's performance on testing data.

\begin{figure}[!tb]
	{\includegraphics[width=3.3in]{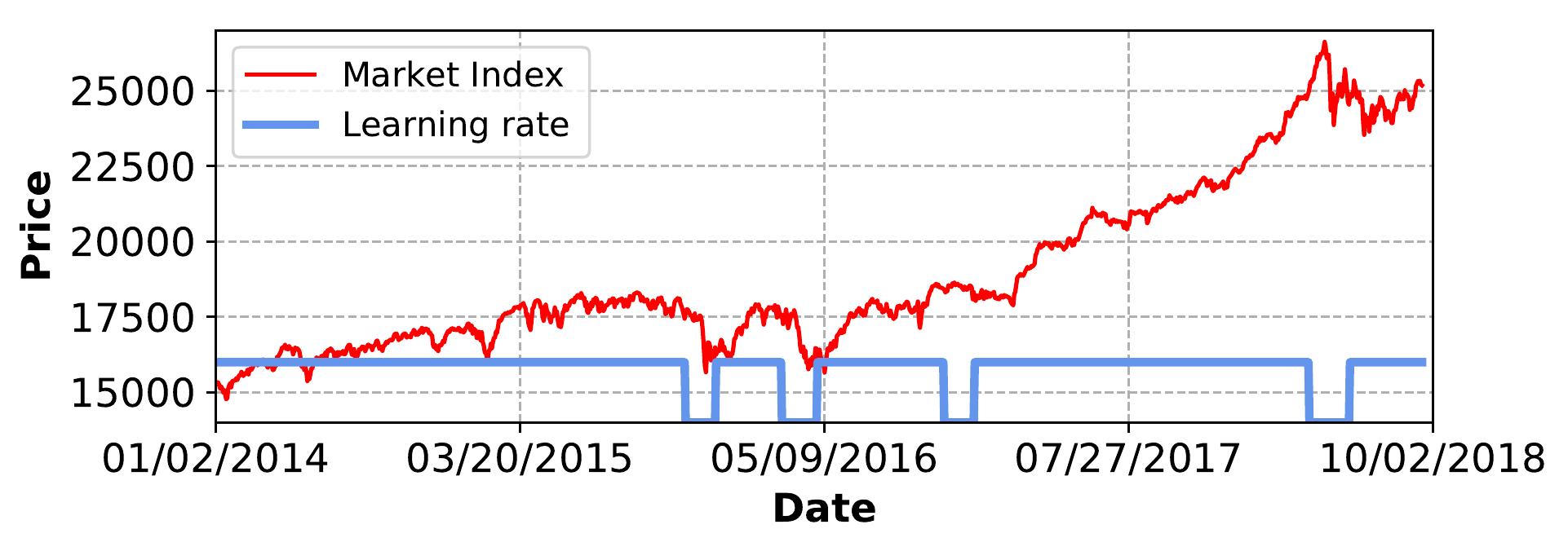}}
	\caption{The market index during the testing data and the corresponding learning rate.}
	\label{figure:market}
\end{figure}

Figure~\ref{figure:market} shows the market index during the testing data and the corresponding learning rate, i.e., we set $\alpha^{+} = 1$ and $\alpha^{-} = 0$. And we set ${\mathcal{N}}^{+}$ as the normal random process and ${\mathcal{N}}^{-}$ as a random process that only generates negative values.

We use the 30 stocks data's daily price to train the adaptive DDPG agent. Then, we run the agent on testing data and compare performance with the vanilla DDPG,  DJIA and the min-variance and mean-variance portfolio allocation strategies. We use final portfolio value, annualized return, annualized standard error and the Sharpe ratio to evaluate the proposed method. Final portfolio value reflects the overall effect of investing in a certain time range. Annualized return is the geometric average amount of money earned by an investment each year over a given time period. Annualized standard error reflects the volatility and shows the robustness of the model.  The Sharpe ratio (the return earned per unit volatility) is used to evaluate the portfolio’s performance~\cite{sharpe1994sharpe}.


\begin{figure*}[!htb]
	\centering
	{\includegraphics[width=7.0in]{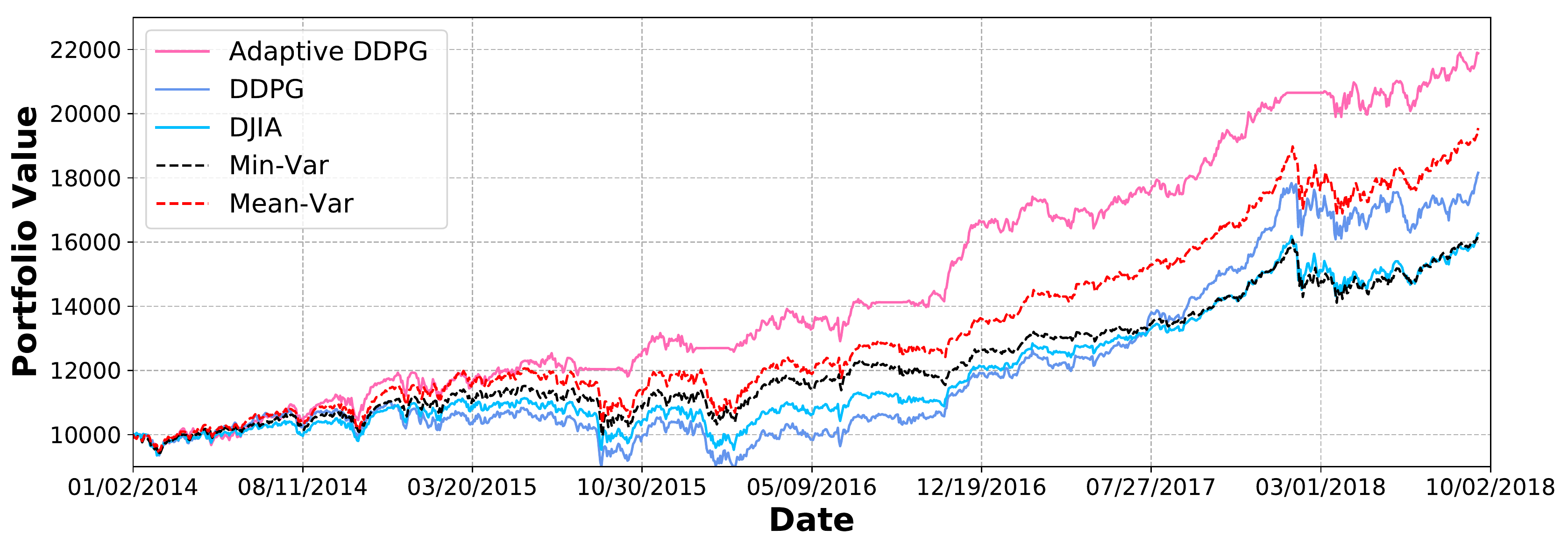}}
	\caption{Portfolio allocation returns of the proposed adaptive DDPG and traditional methods.}
	\label{figure:result}
\end{figure*}

\begin{table*}
\centering
\begin{tabular}{cccccc}  
\toprule
Method  & \textbf{Adaptive DDPG} & DDPG & DJIA & Min-variance & Mean-variance\\
\midrule
Initial value     & \textbf{10,000} & 10,000  & 10,000 & 10,000 & 10,000\\
Final value     & \textbf{21,880} & 18,156 & 16,089 & 16,333 & 19,632 \\
Annualized return      & \textbf{18.84\%} & 14.71\% & 11.36\% & 11.48\% & 15.86\% \\
Annualized Std. error       & \textbf{11.59\%}  & 14.68\% &12.43\% & 11.64\% & 12.70\% \\
Sharpe ratio  & \textbf{1.63}  & 1.01 & 0.91  & 0.99 & 1.25\\
\bottomrule
\end{tabular}
\caption{Trading Performance.}
\label{tab:results}
\end{table*}

\subsection{Performance Results}
Figure~\ref{figure:result} shows that the adaptive DDPG model is significantly better than the vanilla DDPG, the Dow Jones Industrial Average and the traditional portfolio allocation strategies. And we can see that the DDPG strategy is better than the Dow Jones Industrial Average and the traditional portfolio allocation strategies.

As can be seen from Table~\ref{tab:results}, the Adaptive DDPG achieves an annualized rate of return of 18.84\%, which is much higher than the vanilla DDPG of 14.71\%, the Dow Jones Industrial Average of 11.36\%, and the min-variance and mean-variance portfolio allocations respectively of 11.48\% and 15.86\%. The annualized Sharpe ratio of the adaptive DDPG strategy is also higher, indicating that the adaptive DDPG strategy is superior. Therefore, the results show that the adaptive DDPG strategy can effectively develop a matching strategy that is superior to the vinilla DDPG, benchmark Dow Jones industrial average and the traditional portfolio allocation methods.

\section{Conclusions}
In this paper, we propose an Adaptive Deep Deterministic Policy Gradient (Adaptive DDPG) scheme for portfolio allocation task. The Adaptive DDPG incorporates optimistic or pessimistic deep reinforcement learning, which allows the amplitude of the up-date to be different according to the positive or negative prediction errors. Experiment results based on the Dow Jones stocks show that the proposed Adaptive DDPG model can get a better portfolio allocation strategy under different market conditions. Portfolio return results show that the investment return can be significantly improved based on our Adaptive DDPG.

Future work will be interesting to explore more advanced model and deal with larger scale data \cite{burda2018large} to incorporate with price prediction and anomaly detection schemes \cite{Xinyi2019KDD}, and to improve the robustness of machine learning algorithms \cite{yang2018practical}. We also want to do some text analysis, such as extracting text information from real-time news or social networks into the model for analysis \cite{hu2018listening}.

\nocite{langley00}

\bibliography{ourref}
\bibliographystyle{icml2019}



\end{document}